%% file: main.tex
\documentclass[conference]{IEEEtran}
\IEEEoverridecommandlockouts
\usepackage{cite}
\usepackage{mdwlist}
\usepackage{amsmath,amssymb,amsfonts,dsfont,booktabs}
\usepackage{algorithm}
\usepackage[noend]{algpseudocode}
\usepackage{graphicx}
\usepackage{subcaption}
\usepackage{textcomp}
\usepackage{xcolor}
\usepackage{authblk}
\usepackage{lipsum}
\def\BibTeX{{\rm B\kern-.05em{\sc i\kern-.025em b}\kern-.08em
    T\kern-.1667em\lower.7ex\hbox{E}\kern-.125emX}}
    
\makeatletter
\def\BState{\State\hskip-\ALG@thistlm}
\makeatother

\begin{document}
\setlength{\abovedisplayskip}{3pt}
\setlength{\belowdisplayskip}{3pt}

\title{Proactive Islanding of the Power Grid to Mitigate High-Impact Low-Frequency Events}
\author[1]{Shuchismita Biswas}
\author[2]{Emanuel Bernabeu}
\author[2]{David Picarelli\vspace{-0.15in}}

\affil[1]{Department of Electrical and Computer Engineering, Virginia Tech, Blacksburg, VA, USA}
\affil[2]{PJM Interconnection, Audubon, PA, USA\vspace{-0.2in}}

\renewcommand\Authands{ and }

\maketitle

\begin{abstract}

This paper proposes a methodology for enhancing power systems resiliency 
by proactively splitting an interconnected grid into small self-sustaining islands in preparation for extreme events. The idea is to 
posture the system so that cascading outages can be bound within affected areas, preventing the propagation of disturbances to the rest of the system. This mitigation strategy will prove especially useful when advance notification of a threat is available but its nature not well understood. In our method, islands are determined using a constrained hierarchical spectral clustering technique. We further check viability of the resultant islands using steady state AC power flow. Performance of the approach is illustrated using a detailed PSS/E model of the heavily meshed transmission network operated by PJM Interconnection in eastern USA. Representative cases from different seasons show that variations in  power flow patterns influence island configuration. 
\end{abstract}


\input{intro.tex}
\input{method.tex}

\input{results.tex}

\input{conclusion.tex}


\bibliographystyle{./bibliography/IEEEtran}
\bibliography{./bibliography/IEEEabrv,./bibliography/IEEEexample}

\end{document}

%% file: intro.tex
\vspace{-0.08in}
\section{Introduction}\label{sec:intro}

\begin{figure}[t!]
    \centering
    \vspace{-0.08in}
    \includegraphics[clip, trim= 0.6in 2.31in 0.6in 0.18in, width=0.92\columnwidth]{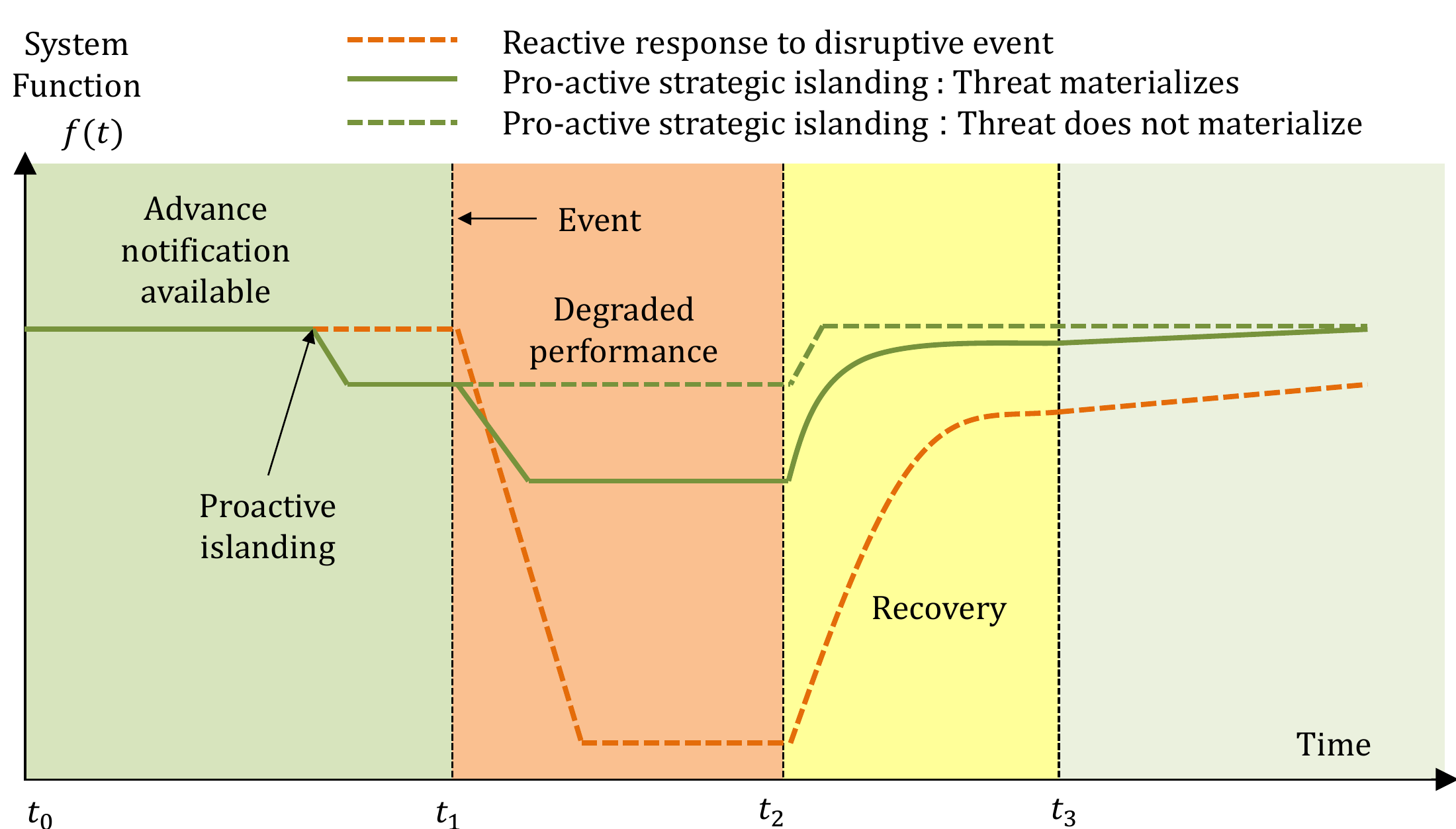}
    \caption{Timeline of proposed proactive islanding strategy}
    \label{fig:timeline}
    \vspace{1pt}
    \includegraphics[clip, trim= 0 1in 0 0,width=0.8\columnwidth]{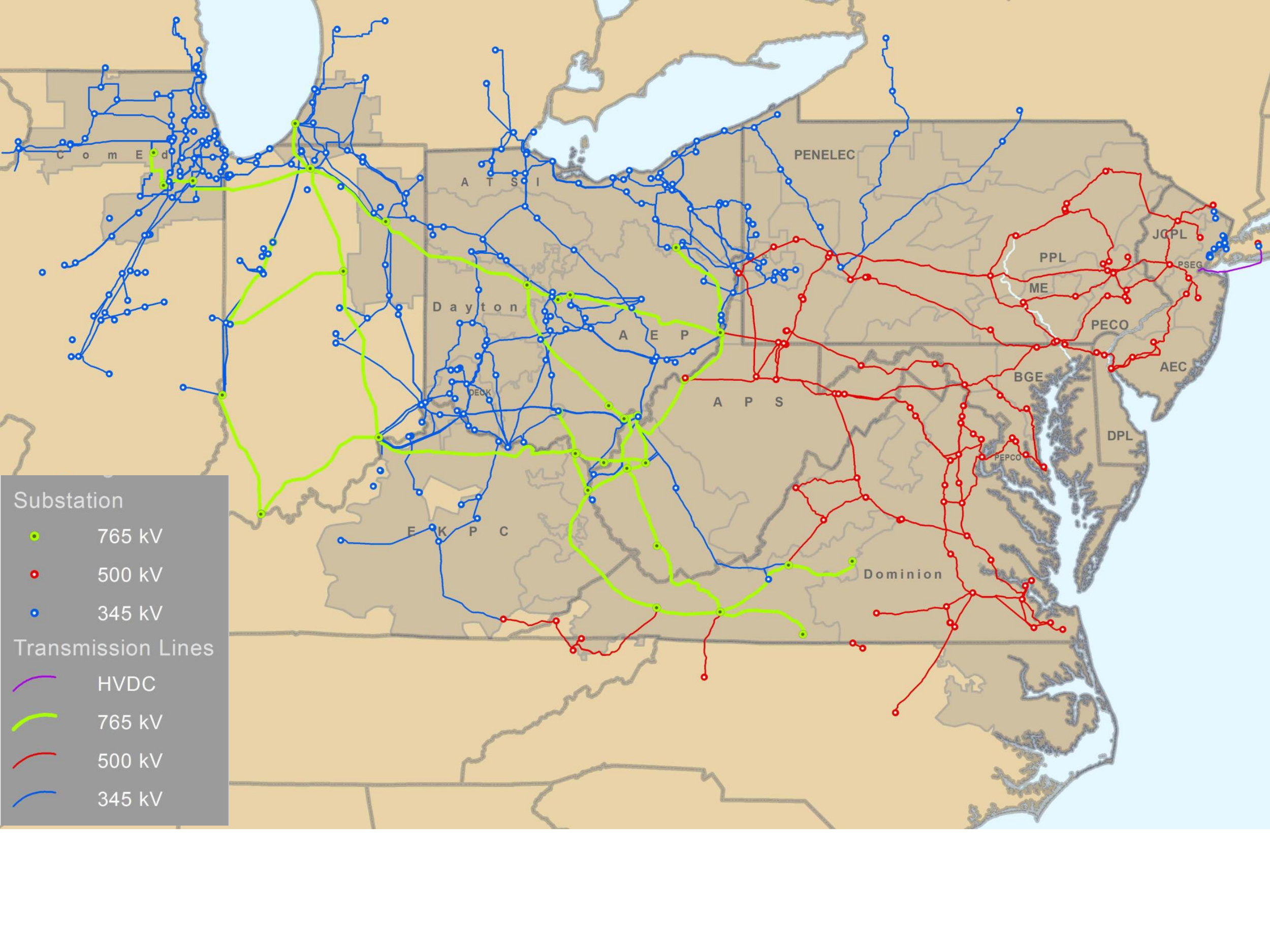}
    \caption{PJM backbone transmission network, showing substations and lines with voltages above 345 kV \cite{pjm_value} }
    \label{fig:PJM}
    \vspace{-0.25in}
\end{figure}

High-Impact Low-Frequency (HILF) events like coordinated cyber, physical or blended attacks, extreme solar weather and high altitude detonation of a nuclear weapon may  cause catastrophic and long-lasting damage to the power grid \cite{HILF, weather_CRS,cybersecurity_CRS,PJM_manual13,meyur}. Although such events are rare, 
the grid should be equipped to mitigate and recover from their effects. 
In the interconnected AC grid, an initial disturbance may cause large-scale outages due to cascading failures \cite{cascading_tree}. The industry deploys intentional controlled islanding (ICI) as a corrective action to arrest cascading events  \cite{Li_controlled_partitioning,tortos,ICI_measurement,ICI_MILP,ICI,ICI_ahad}, but these are \textit{reactive responses} to faults and need to be executed swiftly.  
This paper, in contrast, proposes to proactively partition the grid into self-sustaining islands before the disturbance occurs, if credible intelligence of an imminent threat is available. This idea is illustrated in fig. \ref{fig:timeline}. The y-axis shows system performance as a function of time (load served, reliability etc). As the figure suggests, despite prior intelligence, conventional reactive responses are deployed only when an event occurs. Proactive action may degrade system performance before the event, but will subsequently help in limiting system damage by arresting cascading events, thereby also making recovery easier. Since actions are initiated prior to the event, operators have time to coordinate control actions needed to form stable islands. 

This mitigation approach may prove especially useful when threat intelligence is limited. For example, attacks on the grid may be anticipated, but exact target locations might not be known. All threats may also not materialize. 
Hence, islands created should meet reliability criteria,  
survive for extended time periods and minimize load-shedding. 
Ad hoc operator actions will be needed for continued service, so prevailing operating conditions and other concerns like maintaining system awareness and observability need to be addressed.

Some ideas pertaining to proactively splitting the grid appear in literature. In \cite{panteli_weather}, the authors assess failure probability of components due to a weather contingency and isolate those most likely to fail within an island. This approach is effective only if the event is localized. 
A network splitting strategy is also proposed in \cite{spectral}, but it is not checked if the sub-networks can survive as islands. In this paper, we outline a flexible methodology for determining islands within a highly meshed grid and further show their viability using steady state AC power flow.  Our approach addresses prevalent system conditions, is independent of where the disturbance originates and  can be tailored to the nature of the contingency. The proposed scheme is demonstrated using a PSS/E model of the PJM transmission network in the Eastern Interconnection (EI). System conditions representative of different seasons are considered. 
Due to the vast expanse and geography of the PJM footprint (fig. \ref{fig:PJM}), and the intrinsic dense, highly meshed nature of its infrastructure, PJM has no intuitive `natural' islanding interfaces.
Thus, strategically islanding PJM is a complex task.


%% file: method.tex
\vspace{-0.06in}
\section{Methodology}\label{sec:method}
\begin{figure}
    \centering
    \includegraphics[clip, trim= 0in 5.7in 0in 0in, width=\columnwidth]{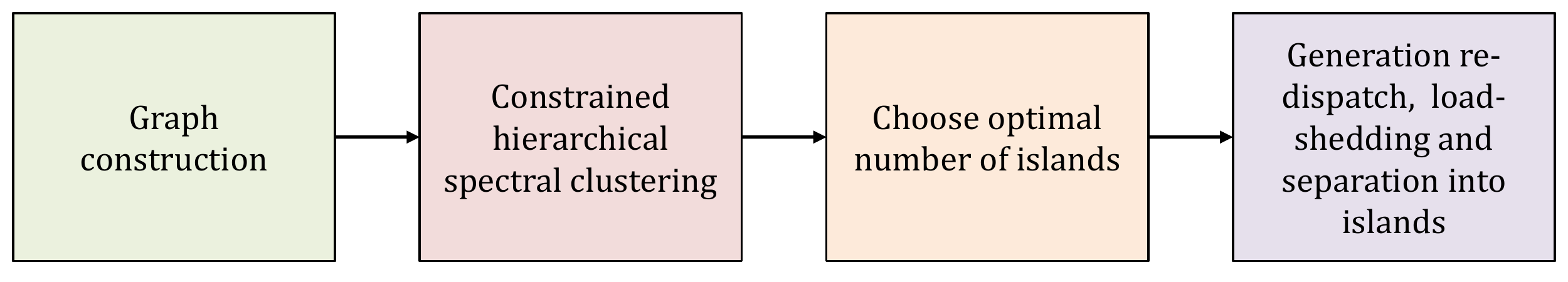}
    \caption{Proposed methodology}
    \label{fig:method}
    \vspace{-0.25in}
\end{figure}

The proposed multi-step approach is described in fig. \ref{fig:method}. The electrical network is represented as a graph and weakly interconnected sub-graphs capable of surviving as islands are identified using a constrained spectral clustering technique. Optimal number of islands is determined based on the nature of the expected contingency and the network needs to be split accordingly.  An overview of each step is provided next.
\vspace{-0.07in}
\subsection{Graph Construction}\label{subsec:graph}

The electrical network is represented by a simple weighted undirected graph $\mathcal{G}=(\mathcal{V},\mathcal{E})$, where $\mathcal{V}$ and $\mathcal{E}$ are the vertex and edge sets  respectively. Each PJM transmission zone is a graph vertex. 
If two zones $i$ and $j$ are electrically connected, there is an edge $e_{i,j}$ between the corresponding vertices $v_i$ and $v_j$. The edge weights $w_{i,j}$ are the absolute values of the apparent power exchange between zones i.e. $w_{i,j}$ is the sum of apparent power flows on all tie-lines connecting zones $i$ and $j$. The higher the weight of an edge, the tighter the coupling between vertices. HVDC tie-lines are excluded from the graph as they do not propagate disturbances. 
  All electrical connections to areas outside PJM are mapped to a single vertex $v_x$. In this paper, zones are anonymized. 

 $\mathcal{G}=(\mathcal{V},\mathcal{E})$ is a simple graph as there are no loops or multiple edges between any pair of vertices. Choosing apparent power as weights ensures that both active and reactive power are considered while determining which zones are tightly coupled. Reactive power balance within islands is essential to maintain healthy voltage profiles \cite{Li_controlled_partitioning}. 
Representing zones as graph vertices will ensure that all equipment within a zone stays in the same island. This serves the following purposes.
\begin{itemize*}
    \item Equipment within each zone is owned and operated by a single transmission owner. Therefore, it is pragmatic if operators within a utility control room do not have to maintain different islands.  
    \item Only tie-line power flow measurements are required for graph construction. Usually, tie lines between different areas are better instrumented. 
\end{itemize*}

Moreover, using zones as vertices instead of individual buses greatly limits the size and order of the graph to be clustered. This, in turn, helps us devise a fast adaptive algorithm for island determination, allowing more time for establishing load-generation balance and coordinating switching actions.  

\vspace{-0.05in}
\subsection{Hierarchical Constrained Spectral Clustering}
\vspace{-0.04in}
ICI can essentially be described as NP-hard searching problems on graphs \cite{tortos}. Spectral clustering, a graph theoretic technique, 
has been proposed as an alternative for solving the minimum power flow disruption islanding problem \cite{tortos,spectral,ICI_ahad}. Here, the central idea is to use the eigenvalues and eigenvectors of Laplacian matrices to find groups (\textit{clusters}) of vertices that are highly connected with each other but weakly connected to vertices in other clusters. A comprehensive background on the clustering methodology and its applicability in power systems is provided in \cite{spectral_clustering_tutorial} and \cite{spectral} respectively. Spectral clustering, however, does not consider generator coherency constraints, which is a critical concern for transient stability, especially when islanding the system after a fault, as generators may go out-of-step or cause undamped oscillations. Since the proposed methodology consists of posturing the system, a controlled sequence of switching actions, re-dispatching, and voltage control can be used to ensure transient stability.


\subsubsection{Graph Laplacian}
A simple weighted undirected graph $\mathcal {G}=(\mathcal{V},\mathcal{E})$ can be described with a weighted adjacency matrix $\mathbf{A}$ and a degree matrix $\mathbf{D}$. Let $N$ be the order of graph $\mathcal{G}$ and $w_{i,j}$ the weight of edge $e_{i,j}$. Then, $\mathbf{A} \in \mathds{R}^{N\times N}$ is a symmetric matrix such that,
\begin{equation} 
  [\mathbf{A}]_{i,j}=\begin{cases}
          w_{i,j},             & \text{if } e_{i,j} \in \mathcal{E}\\
          0,                   & \text{otherwise}
\end{cases}   
\label{eq:adjacency}
\end{equation}

$\mathbf{D}$ is a diagonal matrix with non-negative diagonal entries $d_i$, where $d_i$ is the weighted degree of vertex $v_i$ i.e. the total weight of edges connected to that vertex.
\begin{equation}
    d_i=\sum_{j=1}^N w_{i,j}, \quad \forall i\in\{1,2,\dots, N\}
\end{equation}
Laplacian matrices have been used extensively to study graphs. Two main variants are proposed, the unnormalized Laplacian $\mathbf{L}$ and the normalized Laplacian $\mathbf{L_N}$. It is beneficial to use $\mathbf{L_N}$ for clustering purposes since it is scale-independent.
\begin{eqnarray}
\mathbf{L} =&\mathbf{D}-\mathbf{A}\label{eq:L}\\
\mathbf{L_N} =& \mathbf{D}^{-1/2}\mathbf{L}\mathbf{D}^{-1/2}\label{eq:LN}
\end{eqnarray}
    

\begin{figure*}
\captionsetup{justification=centering}
  \begin{subfigure}[b]{0.28\textwidth}
  \centering
    \includegraphics[clip, trim= 0.3in 0.25in 0.3in 0.25in,width=\linewidth]{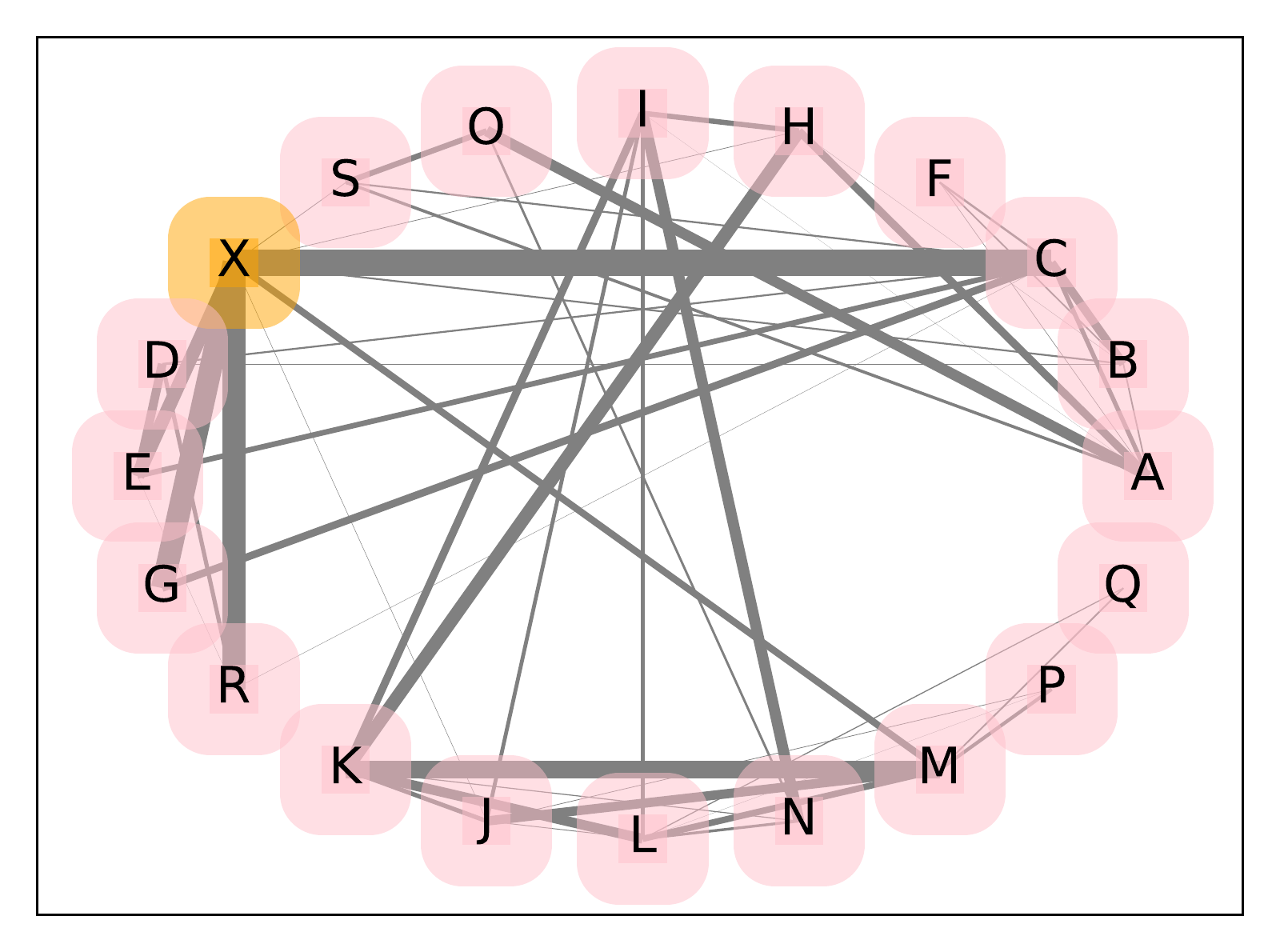}
    \caption{Weighted graph}
    \label{fig:summer_graph}
  \end{subfigure}
 \hfill 
  \begin{subfigure}[b]{0.28\textwidth}
  \centering
    \includegraphics[width=\linewidth]{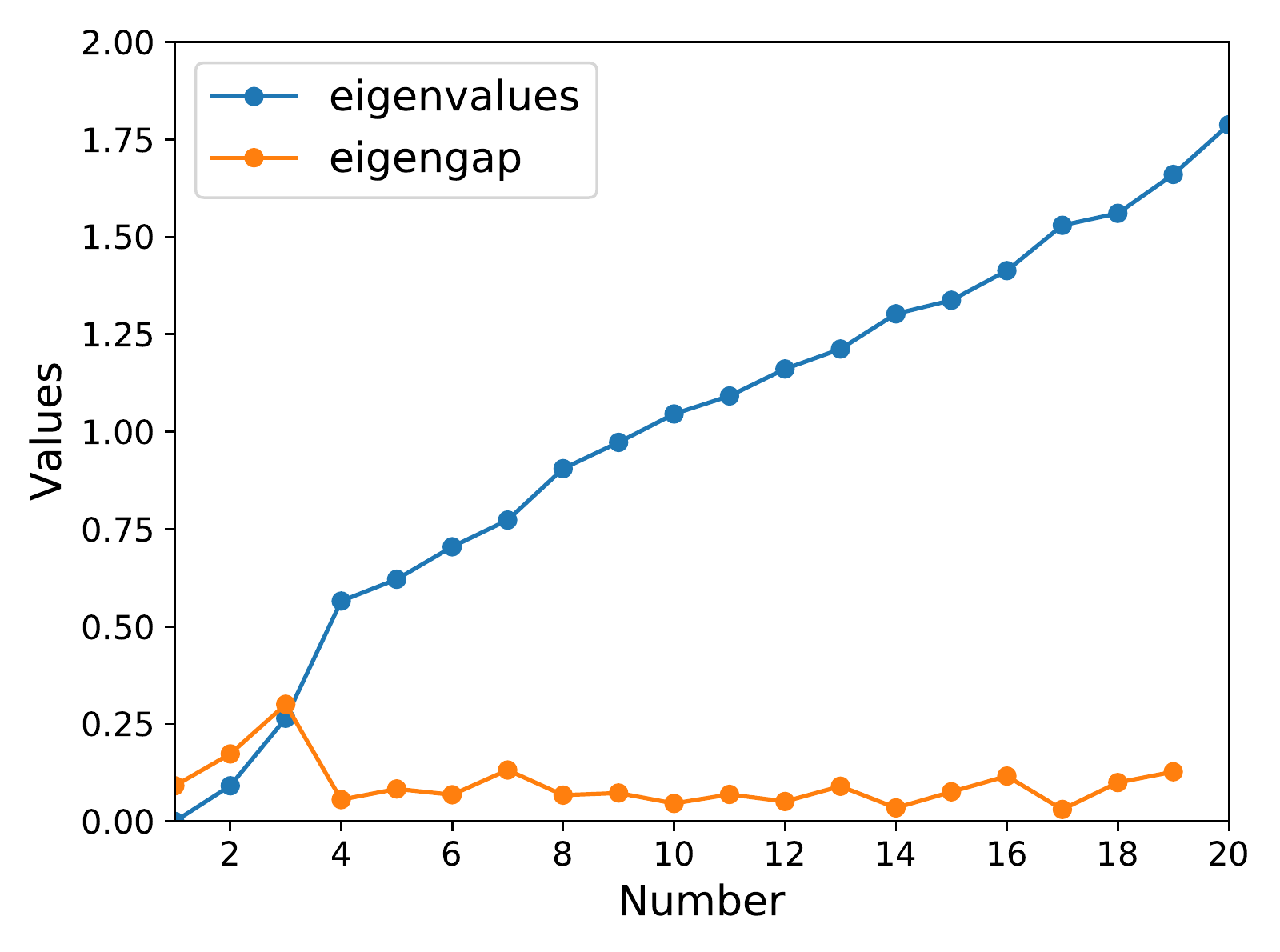}
    \caption{Eigenvalues and eigengaps}
    \label{fig:summer_eigen}
  \end{subfigure}
  \hfill 
 \begin{subfigure}[b]{0.28\textwidth}
 \centering
    \includegraphics[width=\linewidth]{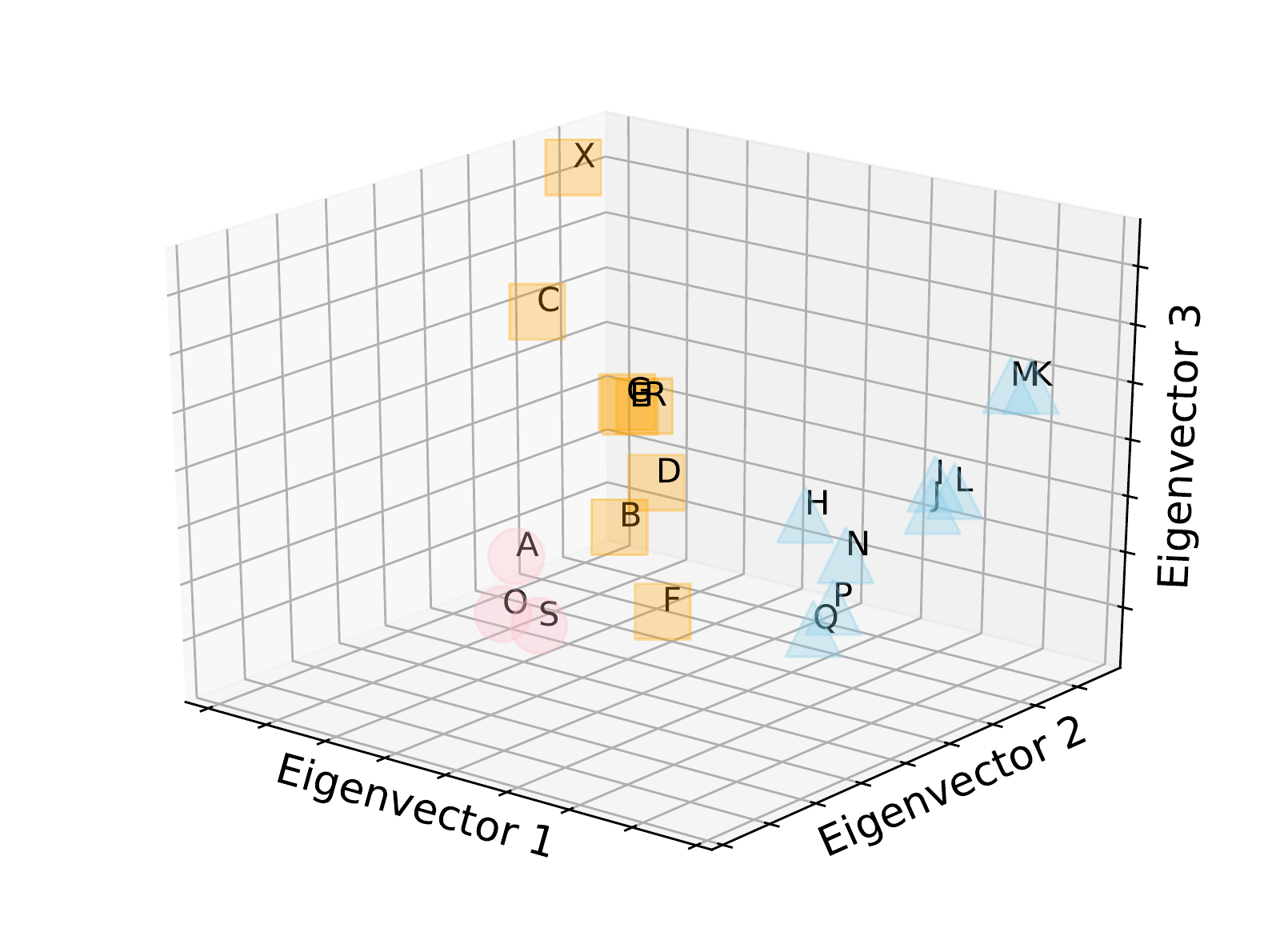}
    \caption{Spectral embedding of graph vertices}
    \label{fig:summer_3d}
  \end{subfigure}
  \hfill 
 \caption{Summer peak load conditions. Eigengap $\gamma_k$ is maximum for $k$=3.}
 \vspace{-0.2in}
  \end{figure*}

\subsubsection{Spectral embedding}
This process refers to representing elements in $\mathcal{V}$ in $k$-dimensional euclidean space $\mathds{R}^k$ using  
the first $k$ eigenvectors of $\mathbf{L}$ or  $\mathbf{L_N}$. Here, $2\leq k \ll N$. One key property of graph Laplacians is that they have $N$ non-negative real eigenvalues
$0\leq\lambda_1\leq\lambda_2\dots\leq\lambda_{N-1}$ \cite{spectral_clustering_tutorial}. First $k$ eigenvectors refer to those corresponding to the $k$ smallest eigenvalues. Ordering the $k$ eigenvectors as columns gives us a  matrix  $\mathbf{X} \in \mathds{R}^{N\times k}$ with rows $\boldsymbol{\chi}_i, i\in \{1,2,\dots,N\}$. Vector $\boldsymbol{\chi}_i$ gives the coordinates  of vertex $v_i$ in $\mathds{R}^k$. Any standard algorithm like k-means may then be used to group these points into clusters. When using $\mathbf{L_N}$, vectors $\boldsymbol{\chi}_i$ must be normalized to length 1 before clustering \cite{spectral_clustering_tutorial,spectral}.

An obvious question here is how to choose $k$. We use the commonly suggested eigengap criterion. Eigengaps are the difference between two consecutive eigenvalues, $\gamma_k=\lambda_{k+1}-\lambda_k$. A high value of $\gamma_k$ suggests that the graph maybe decomposed into at least $k$ clusters and this will be revealed with spectral embedding in $k$-dimension. 

\subsubsection{Constrained hierarchical clustering}
\setlength{\textfloatsep}{0pt}
\begin{algorithm}[t]
\caption{Hierarchical Constrained Spectral Clustering}\label{algo}
\begin{algorithmic}[1]
\State \textbf{Normalized Laplacian: }Compute $\mathbf{L_N}$, as per eq. \eqref{eq:adjacency}-\eqref{eq:LN}.
\State  \textbf{Eigenvalue Decomposition: }Compute eigenvalues and eigenvectors of $\mathbf{L_N}$. \State \textbf{Spectral Dimension: }Sort the eigenvalues of $\mathbf{L_N}$ in ascending order,
       $0\leq \lambda_1\leq\lambda_2\dots\leq\lambda_N$. Choose $2\leq k \ll N$ such that eigengap  $\gamma_k$  is high.
\State \textbf{Spectral $\mathbf{k}$-embedding: } Construct matrix $\mathbf{X}$ with first $k$ eigenvectors of $\mathbf{L_N}$ as its columns. Normalize the columns of $\mathbf{X}$ to length 1. The $i$-th row of $\mathbf{X}$, $\boldsymbol{\chi}_i$ represents the coordinates of vertex $v_i$ in $\mathds{R}^k$.
\State \textbf{Constrained Hierarchical Agglomerative Clustering: } Agglomerative hierarchical clustering of points represented by  vectors $\boldsymbol{\chi}_i, i\in \{1,\dots N\}$ with additional connectivity constraint.
\end{algorithmic}
\vspace{-3pt}
\end{algorithm}

Several limitations exist when applying spectral clustering to the controlled islanding problem. First, if $k>2$, an additional k-means step needs to be performed to identify clusters. This has several drawbacks like- a) number of clusters needs to  be specified a-priori, and b) clustering results depend on the initial choice of centroids. Second, when projecting graph vertices into $\mathds{R}^k$, edge information is ignored. Therefore, there maybe some points which are close in euclidean space but do not have an edge connecting them in the original graph.

The first limitation is overcome using agglomerative  hierarchical clustering, similar to the approach in \cite{spectral}. In this method, at the initial step, every point is considered as an individual cluster. Next, closest clusters according to some distance metric are merged together. This process is repeated until all points are merged into a single cluster. This clustering `hierarchy' can be encoded into a tree-like structure called dendrogram. By `cutting' the dendrogram at different levels, different numbers of clusters can be obtained. We use the criteria outlined in section \ref{subsec:no_islands} in conjunction with the dendrogram to decide the number of islands. 
The ward distance metric is used to determine distance between points.

The second shortcoming is dealt with by imposing a connectivity constraint, 
i.e. clusters are merged only if there is an edge connecting them in the original graph. The complete process is summarized in algorithm \ref{algo}. Here, step 2 is computationally the most expensive, and is at most cubic in $N$ \cite{spectral}.

\vspace{-0.1in}
\subsection{Optimal number of islands} \label{subsec:no_islands}
The dendrogram 
describes the grouping of zones within the electrical network, and by cutting it at different levels, any $r$ number of islands may be determined. How to choose $r$ is a risk-management decision and must consider several factors: 
\begin{itemize*}
    \item What is the nature of the contingency? If damage is expected to be localized, a small number of islands is advisable. For a coordinated attack at multiple points of the network, a larger number of islands might be useful.
    \item Are the islands self-sustainable? We favor solutions that minimize load generation-imbalance and operational violations like line overloading, undervoltages etc.
    \item How many switching operations would be required to split the network? Evidently, solutions that require fewer number of line disconnections are easier to realize. 
\end{itemize*}
\vspace{-0.07in}
\subsection{Separation into islands}
Island creation involves switching a large number of transmission lines. For instance, disconnecting PJM from the Eastern Interconnect (EI) would involve switching 212 lines \cite{pjm_value}. We propose that the splitting is carried out in two steps:
 \subsubsection{Redispatch generation and load shedding within islands}This step will establish load-generation balance within islands and further reduce power flow on tie-lines to be disconnected. Thus, when these lines are switched, there are high chances that generators in islands will stay synchronized and operational violations will be minimized. 
 \subsubsection{Disconnecting transmission lines} Formulating an exact sequence of actions to split the grid into islands is a complicated problem that is not studied in this paper. A review of transmission line switching strategies is provided in \cite{optimal_switching}. We validate the performance of islands looking at steady state AC power flow solutions in PSS/E. Dynamic simulation of the island creation process is not shown.

%% file: results.tex
\section {Simulation and Results}\label{sec:results}

We demonstrate the proposed islanding methodology using PSS/E model of the PJM network. Three different cases are studied, representative of summer peak, winter peak and spring light load conditions. In these cases, load within the PJM network are 163.6 GW, 140.9 GW and 81.5 GW respectively. Counter-intuitively, during light load conditions, inter-area tie-line flow is higher than peak load periods. This is because most generators need to be in service to meet peak load. However, during light load, expensive generators may be turned off to keep energy prices low and cheaper generation units may be located away from load centers. Since studying PJM in isolation is not representative of actual operating conditions, the PSS/E simulation model includes full representations of systems in the EI. The model has more than 158,000 buses, 20,437 of which are within PJM. The PJM network has 1617 generator buses, 944 fixed shunts and 2084 variable shunts.

An interesting question to consider is: should some PJM zones remain connected to the EI when splitting into islands?  In this paper, we consider that certain zones within PJM stay connected to the EI. The methodology can be easily extended to the second case, and would be equivalent to removing vertex `X' from the network graph.
\vspace{-0.05in}
\subsection{Case Studies}
\begin{figure}[t]
    \centering
    \includegraphics[clip, trim =0 0.17in 0 0 ,width=0.7\columnwidth]{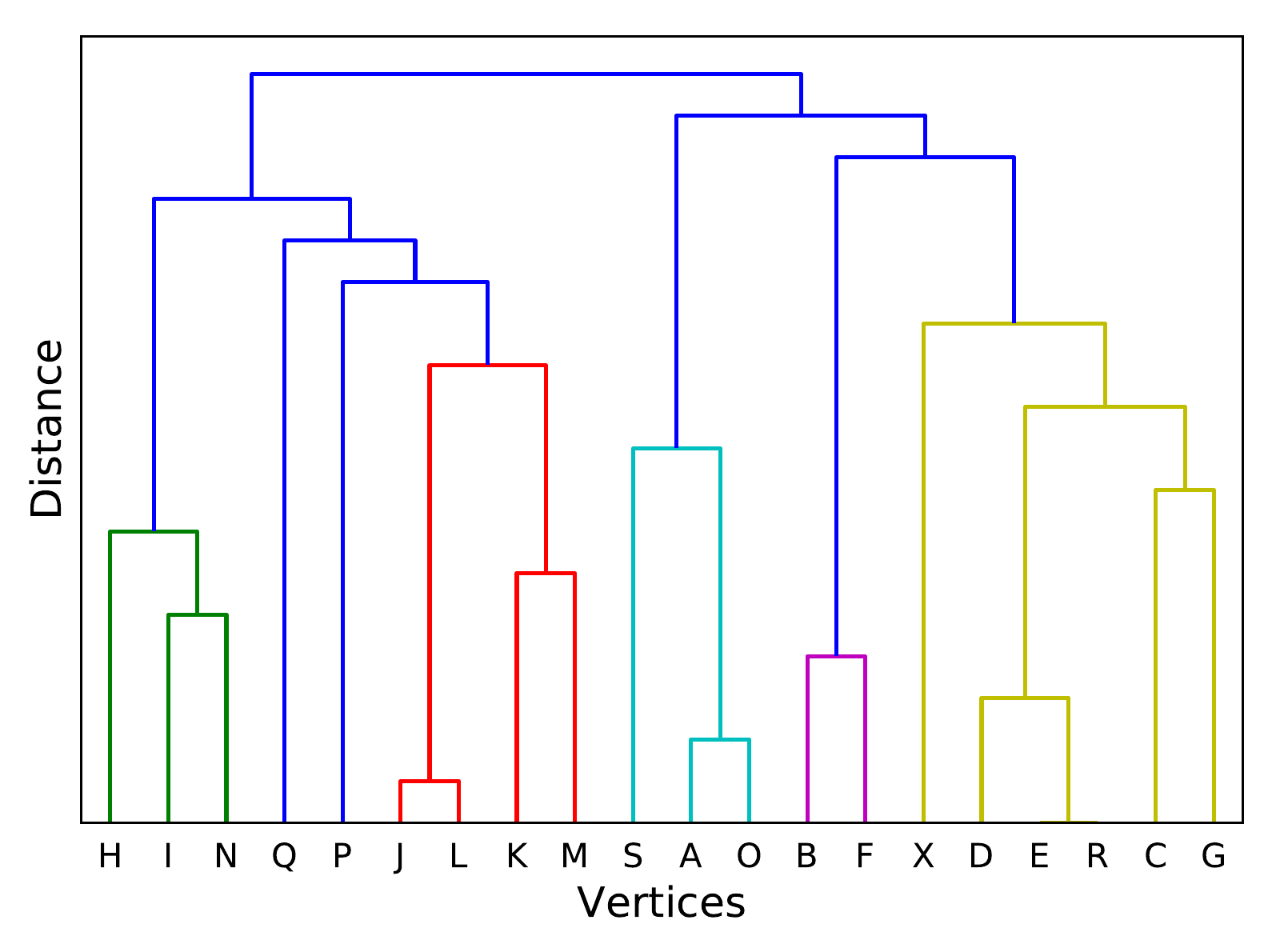}
    \caption{Dendrogram for summer peak load conditions}
    \label{fig:summer_dendro}
    \begin{minipage}{0.45\columnwidth}
    \centering
    \includegraphics[clip, trim =0 0.1in 0 0 ,width=\linewidth]{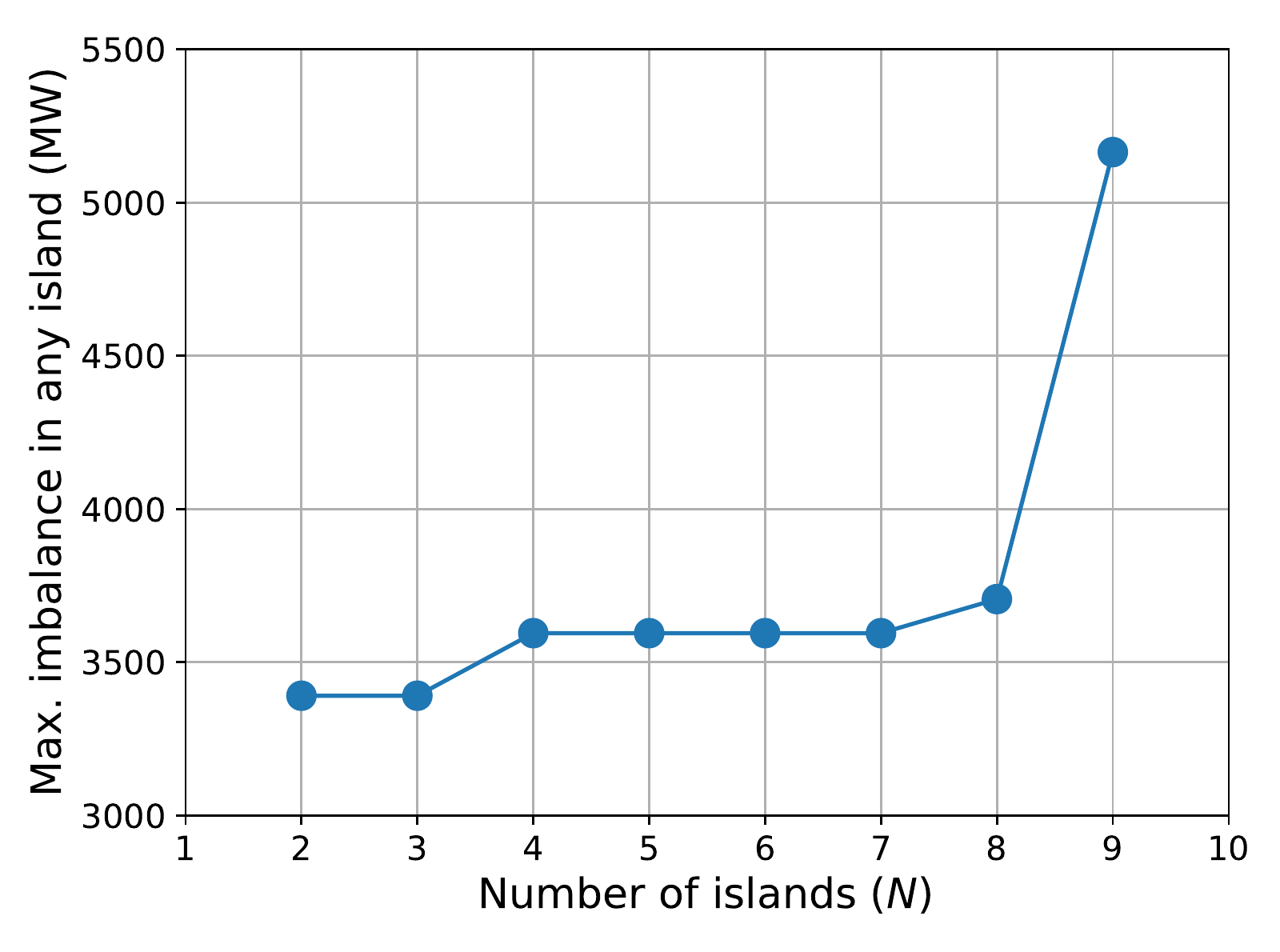}
    \captionsetup{justification=centering}
    \caption{Maximum load generation imbalance}
    \label{fig:summer_balance}
    \end{minipage}
    \hspace{0.06 \columnwidth}
    \begin{minipage}{0.45\columnwidth}
    \includegraphics[clip, trim =0 0.1in 0 0 ,width=\linewidth]{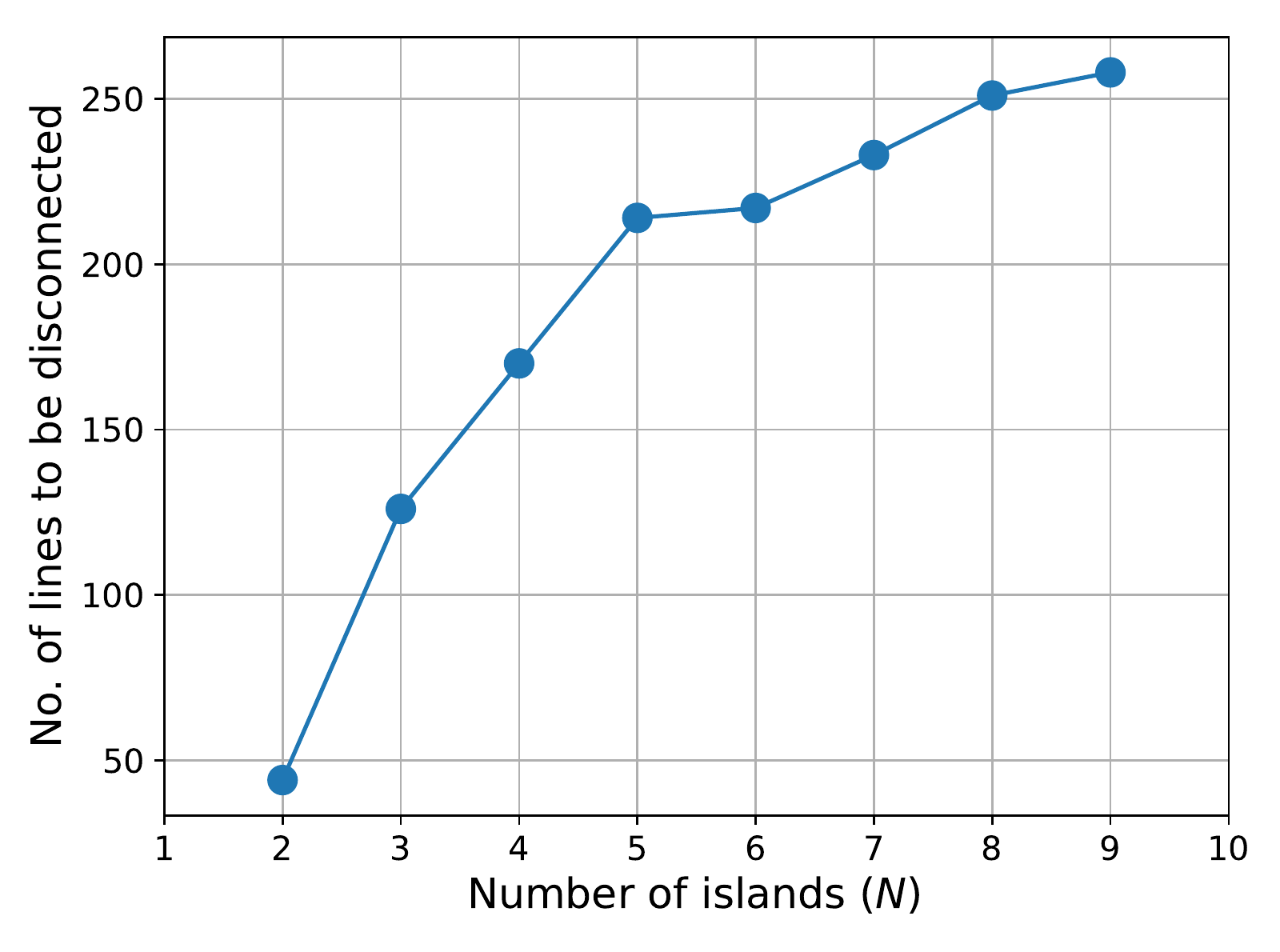}
    \captionsetup{justification=centering}
    \caption{Number of tie-lines to be disconnected}
    \label{fig:summer_switching}
    \end{minipage}
\end{figure}

\vspace{-0.05in}
Let us first consider the summer peak case example.
Fig. \ref{fig:summer_graph} shows the network graph. Edge thicknesses are proportional to their weights. The network outside PJM is mapped onto vertex `X'. Eigenvalues and eigengaps are computed for $\mathbf{L_N}$ (fig. \ref{fig:summer_eigen}). Since eigengap $\gamma_k$ is maximum for $k=3$, spectral embedding is done in three dimensions. This is shown in fig. \ref{fig:summer_3d}. Hierarchical clustering with connectivity constraints yields the dendrogram shown in fig. \ref{fig:summer_dendro}. Grouping into three clusters according to the dendrogram is also shown in fig. \ref{fig:summer_3d}.

\begin{figure}[t]
\begin{minipage}{0.45\columnwidth}
    \centering
    \includegraphics[clip, trim=0 0.5in 0 0, width=\linewidth]{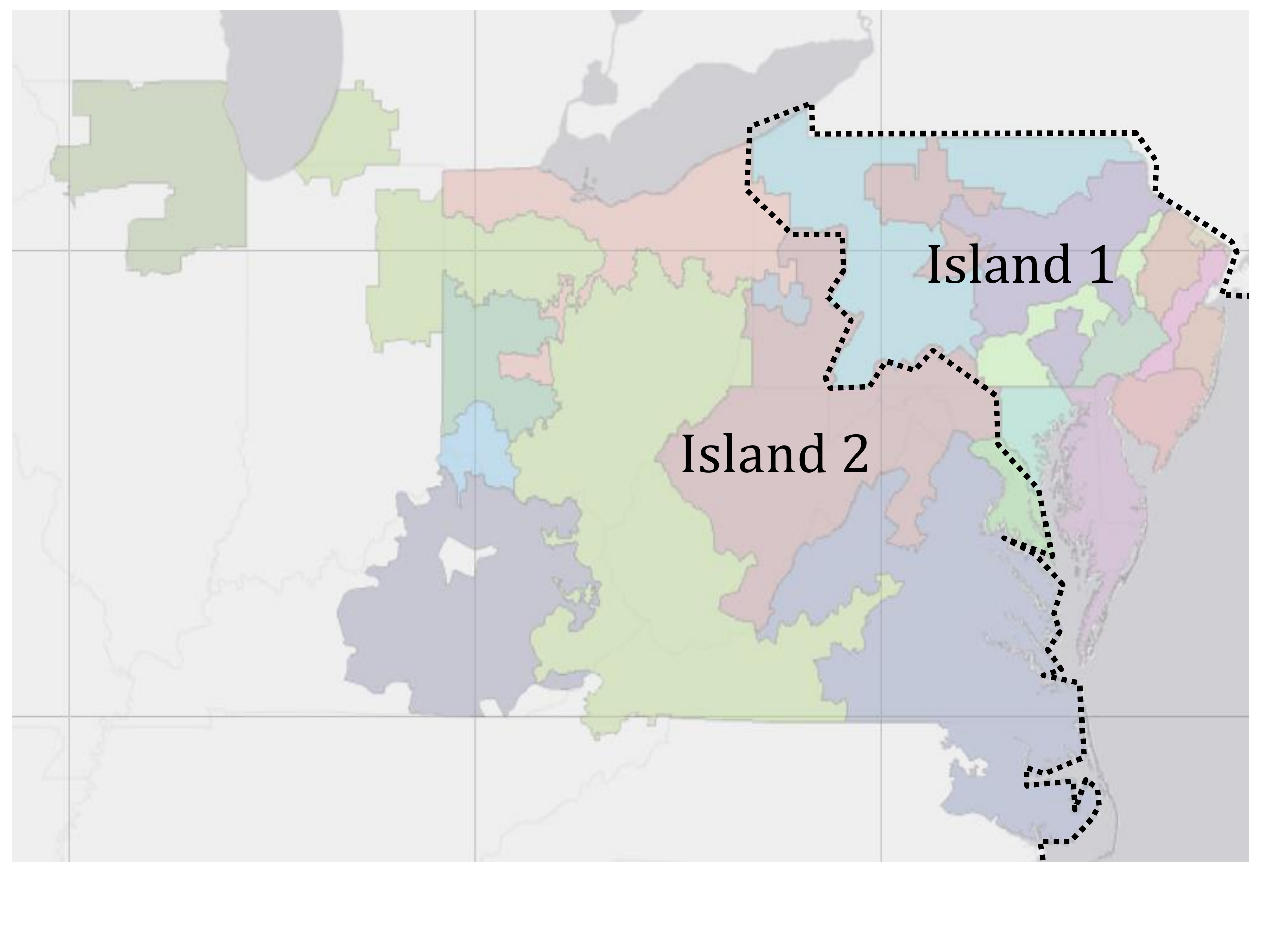}
    \captionsetup{justification=centering}
    \caption{Two islands for summer peak conditions}
    \label{fig:summer2}
    \end{minipage}
    \hspace{0.06 \columnwidth}
    \begin{minipage}{0.45\columnwidth}
    \includegraphics[clip, trim=0 0.5in 0 0, width=\linewidth]{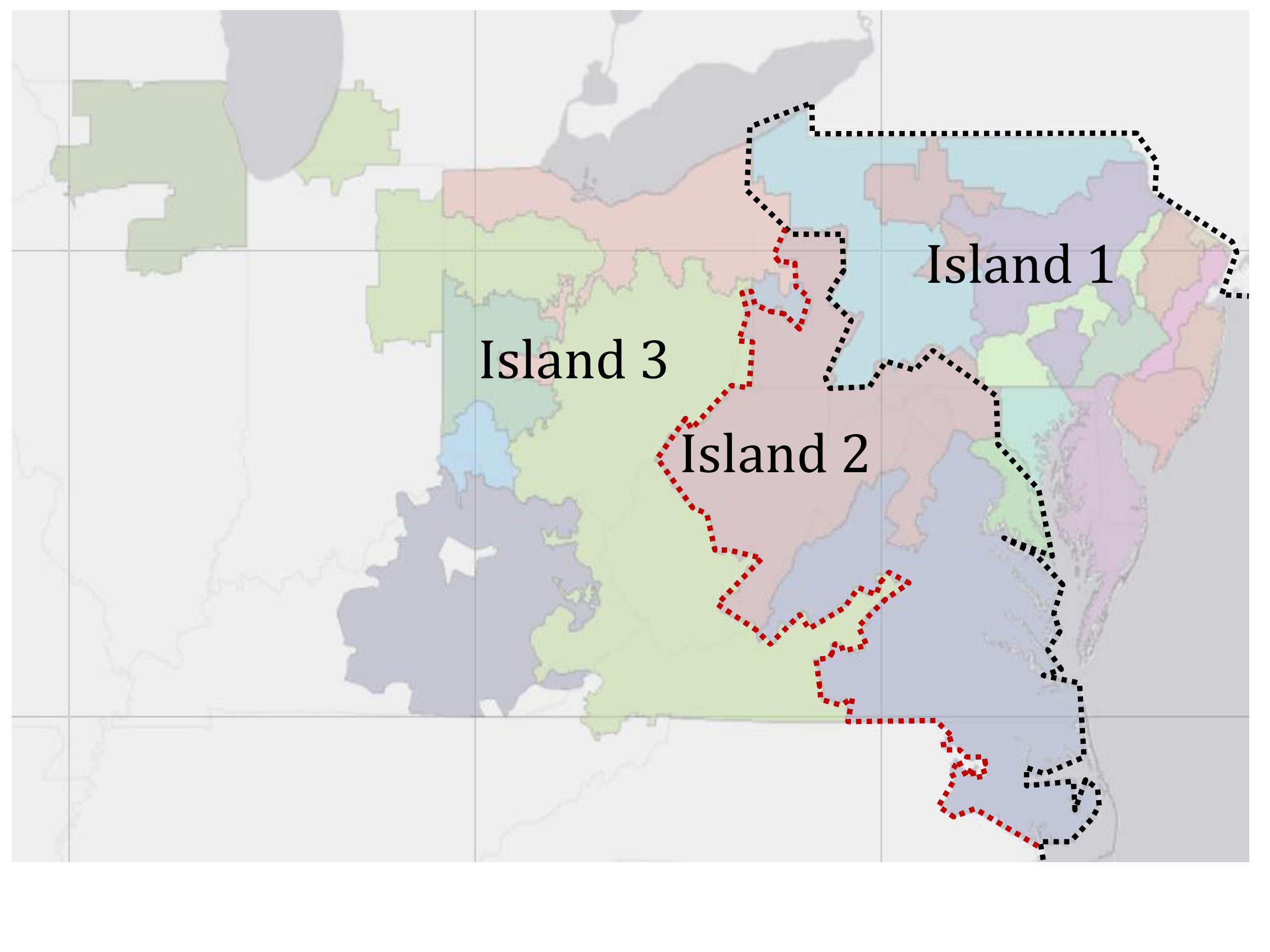}
    \captionsetup{justification=centering}
    \caption{Three islands for summer peak conditions}
    \label{fig:summer3}
    \end{minipage}

    \centering
    \begin{minipage}{0.45\columnwidth}
    \includegraphics[width=\linewidth]{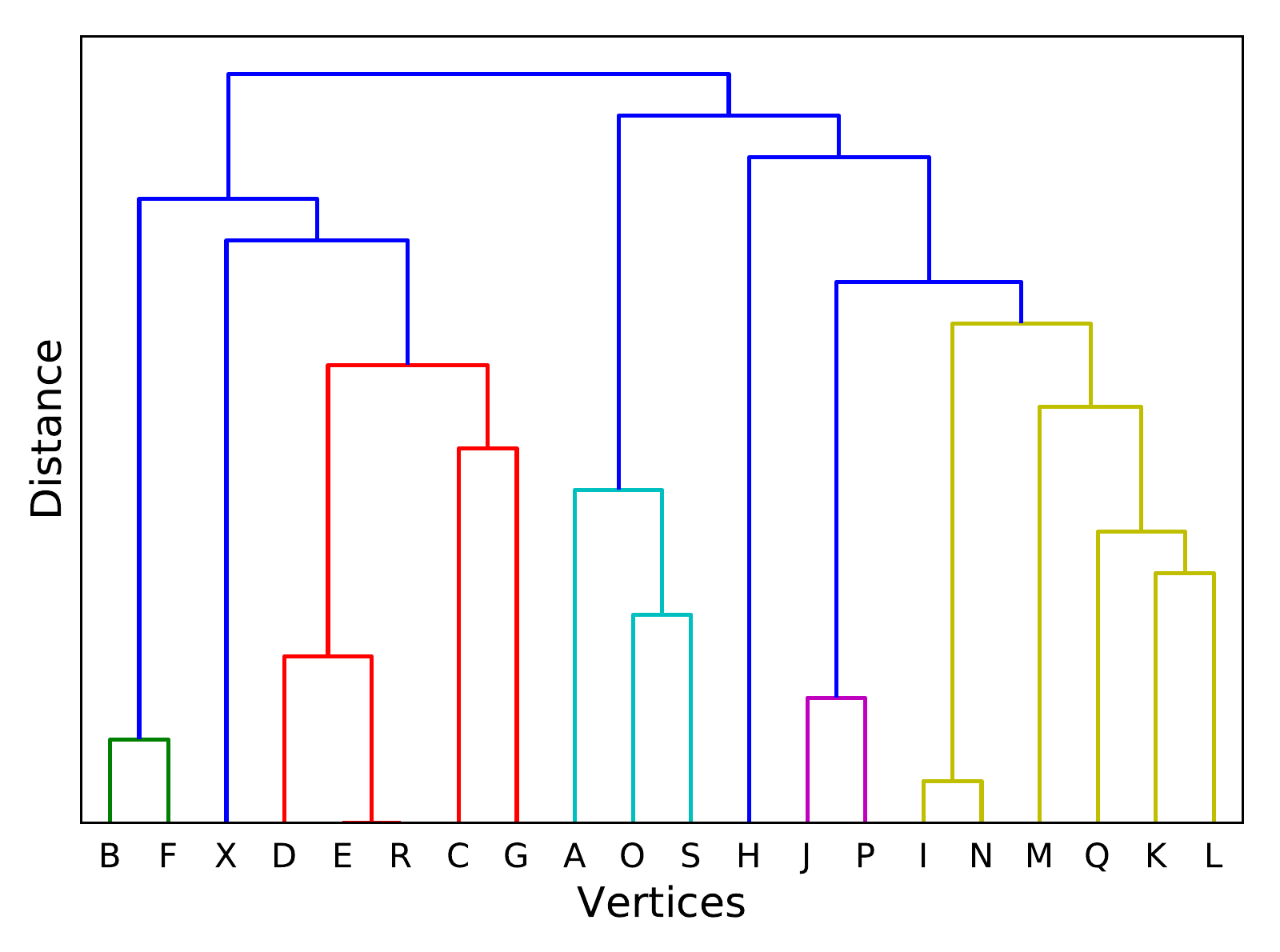}
     \captionsetup{justification=centering}
    \caption{Dendrogram: Winter peak load }
    \label{fig:win_den}
    \end{minipage}
    \hspace{0.06 \columnwidth}
    \begin{minipage}{0.45\columnwidth}
     \includegraphics[width=\linewidth]{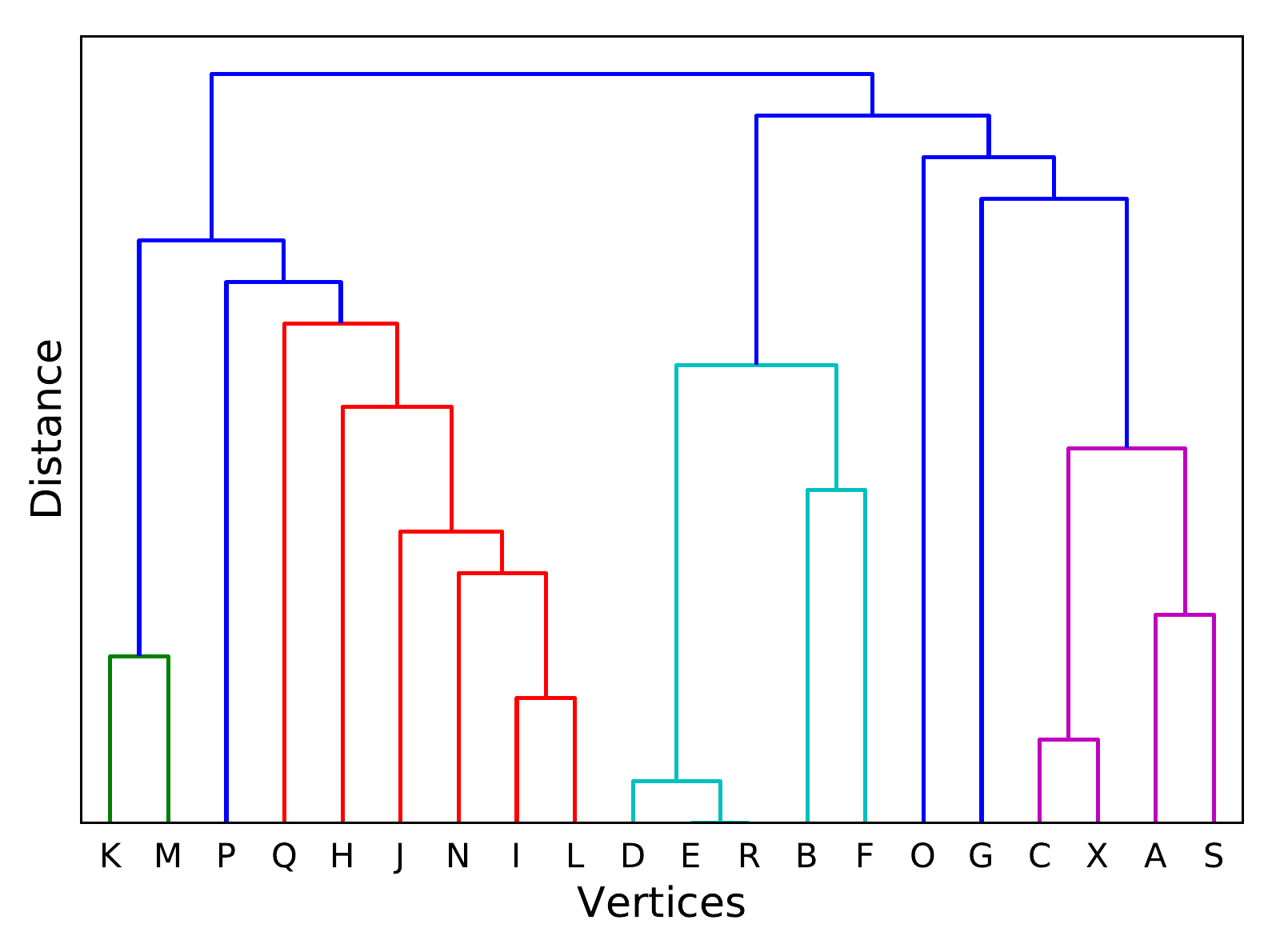}
      \captionsetup{justification=centering}
    \caption{Dendrogram: Spring light load}
    \label{fig:spring_den}
    \end{minipage}
    \vspace{-0.2in}
\end{figure}

\begin{figure}
    \centering
\begin{minipage}{0.45\columnwidth}
    \centering
    \includegraphics[clip, trim=0 0.5in 0 0, width=\linewidth]{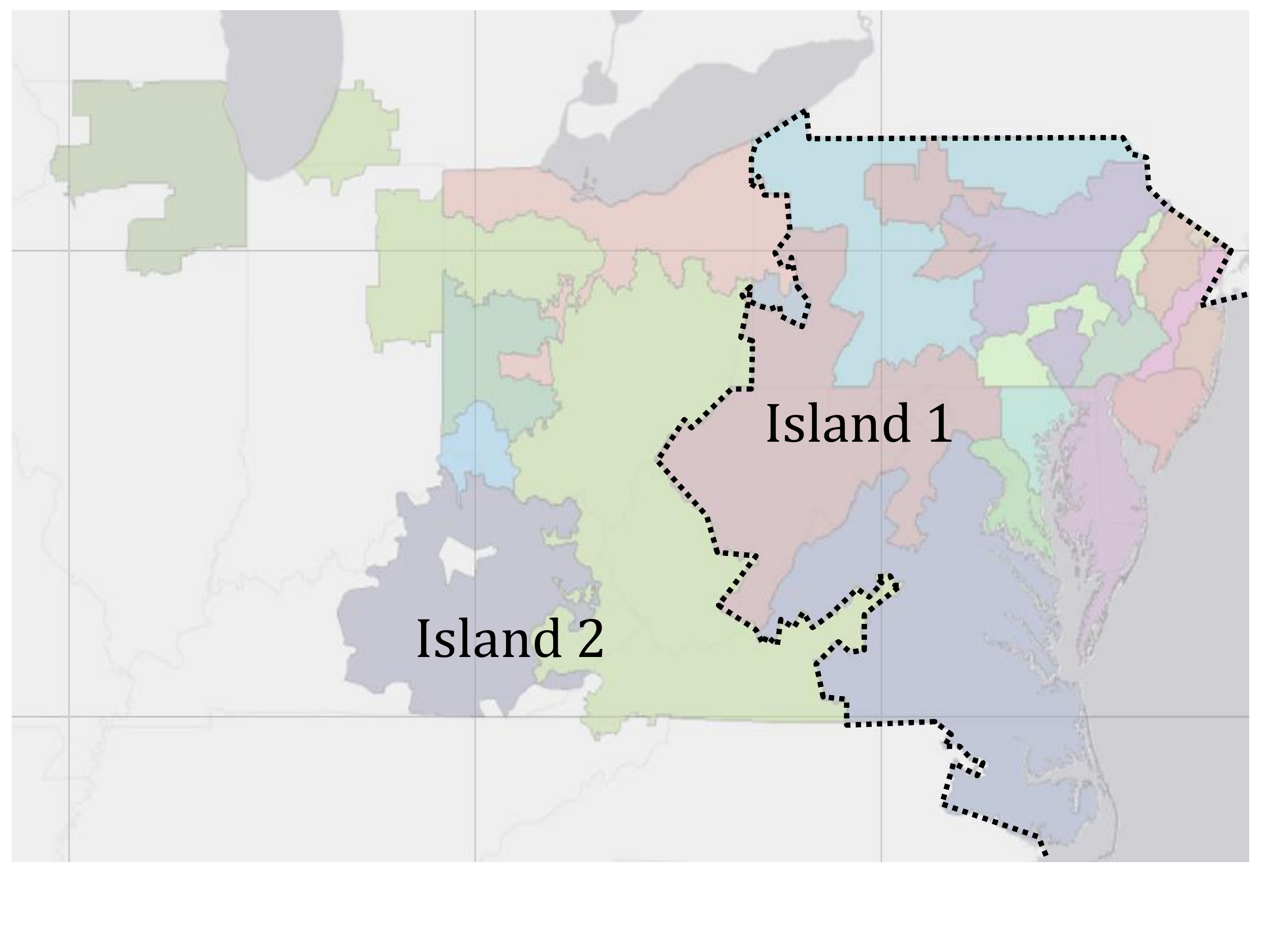}
     \captionsetup{justification=centering}
    \caption{Two islands for winter peak conditions}
    \label{fig:winter2}
    \end{minipage}
    \hspace{0.06 \columnwidth}
    \begin{minipage}{0.45\columnwidth}
    \includegraphics[clip, trim=0 0.5in 0 0, width=\linewidth]{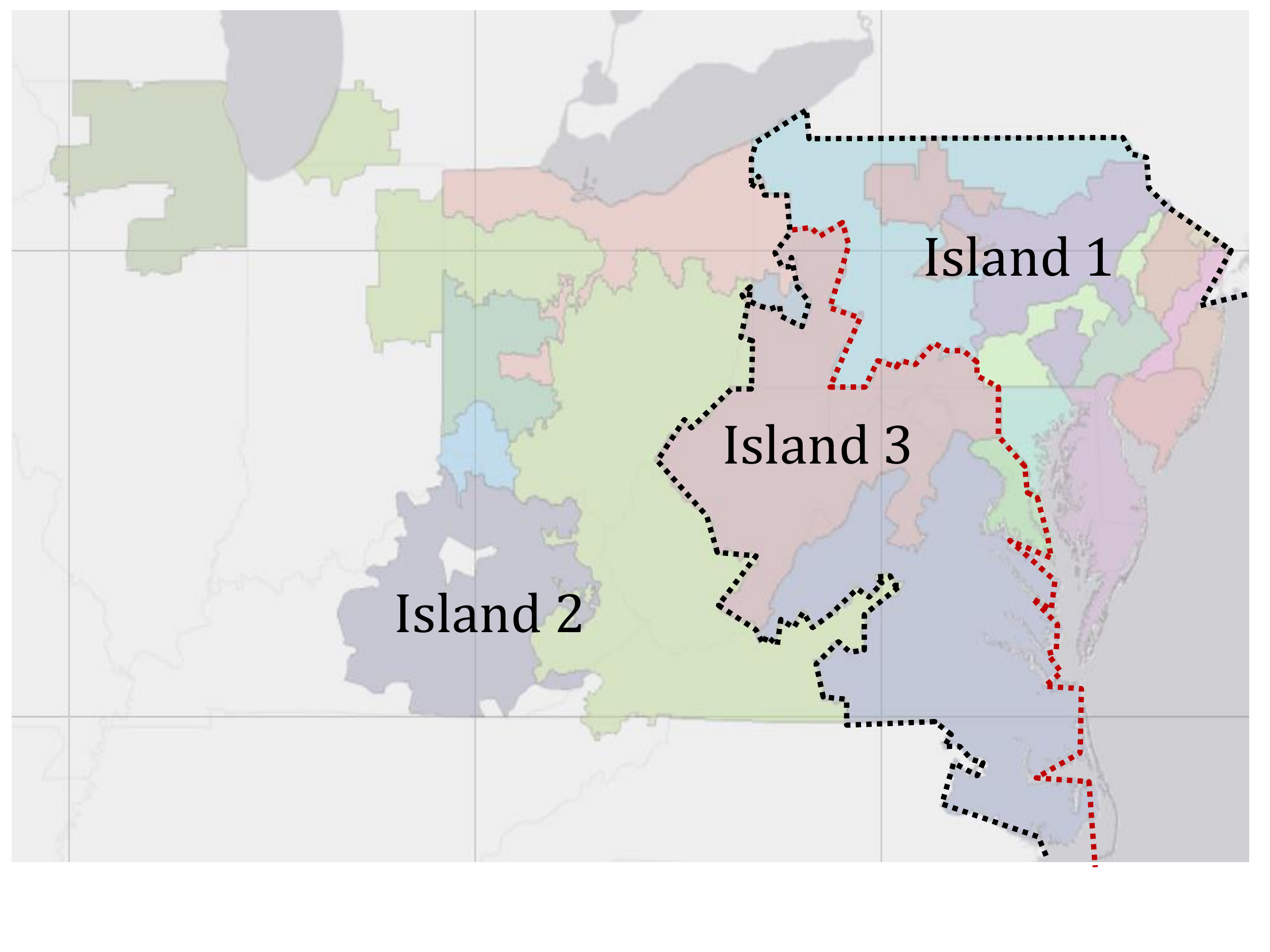}
     \captionsetup{justification=centering}
    \caption{Three islands for winter peak conditions}
    \label{fig:winter3}
    \end{minipage}
\begin{minipage}{0.45\columnwidth}
    \centering
    \includegraphics[clip, trim=0 0.5in 0 0, width=\linewidth]{figures/summer2.pdf}
     \captionsetup{justification=centering}
    \caption{Two islands for spring light load conditions}
    \label{fig:sll2}
    \end{minipage}
    \hspace{0.06 \columnwidth}
    \begin{minipage}{0.45\columnwidth}
    \includegraphics[clip, trim=0 0.5in 0 0, width=\linewidth]{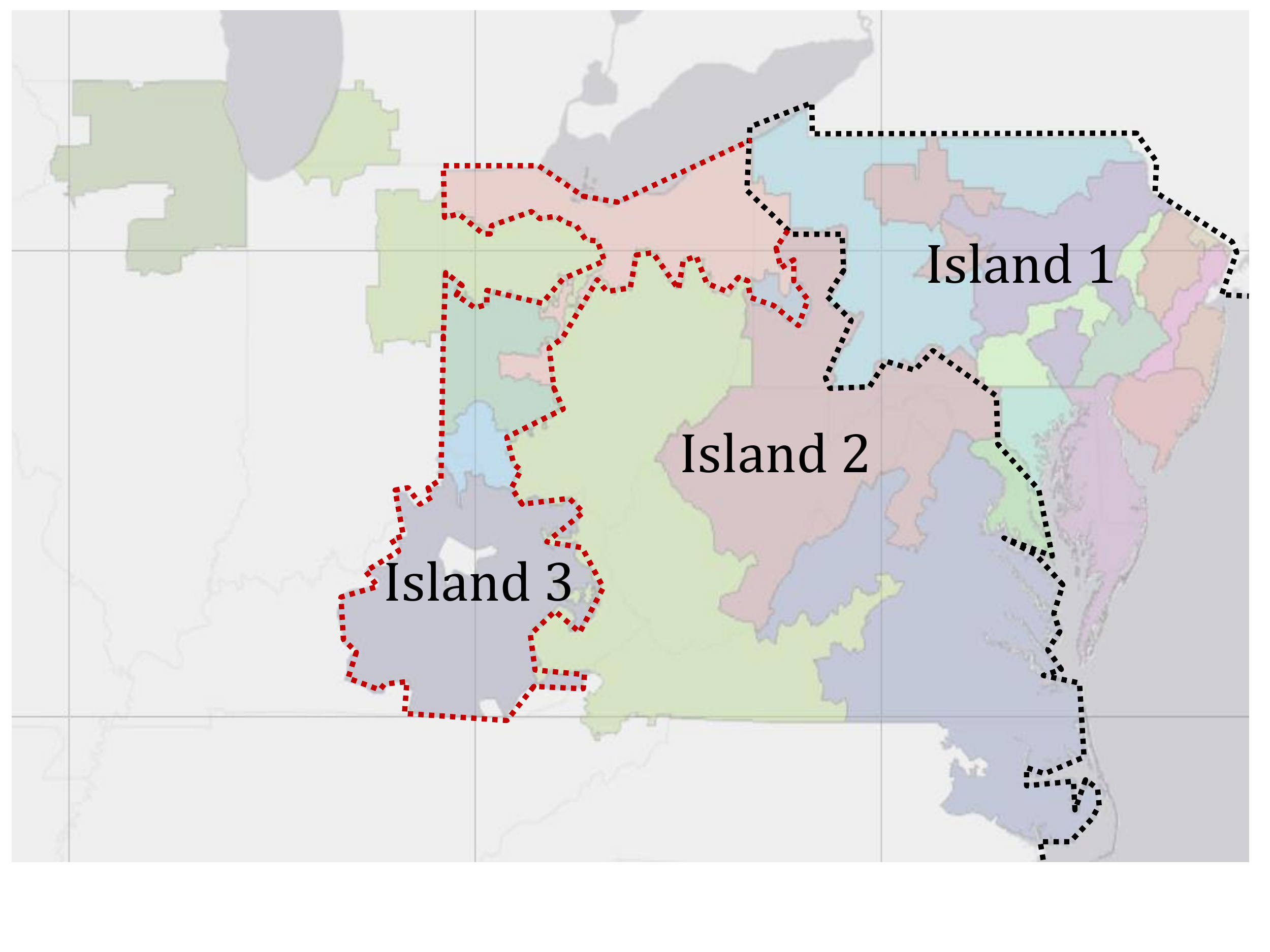}
     \captionsetup{justification=centering}
    \caption{Three islands for spring light load conditions}
    \label{fig:sll3}
    \end{minipage}
\end{figure}

If the number of islands is varied from two to nine, maximum load-generation imbalance expected in any island would be as shown in fig. \ref{fig:summer_balance}. Number of tie-lines to be switched for creating $r$ islands is shown in fig. \ref{fig:summer_switching}. We see that splitting the grid into three islands from two does not increase the maximum expected imbalance, but needs 82 additional switching operations. Hence, the number of islands needed must be decided on a case by case basis. The approximate geographical boundaries of islands are shown in fig. \ref{fig:summer2} and \ref{fig:summer3}.  Using the same methodology, the dendrograms shown in fig. \ref{fig:win_den} and \ref{fig:spring_den} are  obtained for winter peak and spring light load conditions. Approximate island boundaries are shown in fig. \ref{fig:winter2}-\ref{fig:sll3}.  It is evident that island configuration changes with different seasons.

\begin{table}[b]
    \caption{Clustering performance metric $p$}
    \centering
    \begin{tabular}{cccc}
    \toprule
         & \textbf{Summer peak} & \textbf{Winter peak} & \textbf{Spring light load} \\
         \midrule
         \textbf{Two clusters}
         & 0.0702 & 0.0955 & 0.0958 \\
         \textbf{Three clusters}
         & 0.1081 & 0.1766 & 0.2495\\
         \bottomrule
    \end{tabular}
    \label{tab:p}
\end{table}  
\begin{table*}[t]
\centering
\captionsetup{justification=centering}
\caption{Electrical performance when the network is divided into two islands}
\begin{tabular}{ccccccc}
\toprule
 &  \multicolumn{2}{c}{\textbf{Summer Peak}} & \multicolumn{2}{c}{\textbf{Winter Peak}}  & \multicolumn{2}{c}{\textbf{Light Load}}\\
\midrule
{}   & Island 1 & Island 2 & Island 1 & Island 2 & Island 1 & Island 2   \\
\textbf{Gen. redispatch (MW)}   &  -3341.4 & 3033.8   & 1218.9  & -1430.3 & 1313.8 & -1348.1\\
\textbf{Load-shedding (MW}) & 0 & 0 & 0 & 0 & 0 & 0\\
\textbf{Min. bus voltage (p.u.)} & 0.97 & 0.96 & 0.97 & 0.96 & 0.96 & 0.96 \\
\bottomrule
\end{tabular}
\label{tab:2island}
\vspace{8pt}
\centering
\captionsetup{justification=centering}
\caption{Electrical performance when the network is divided into three islands}
\begin{tabular}{cccccccccc}
\toprule
 &  \multicolumn{3}{c}{\textbf{Summer Peak}} & \multicolumn{3}{c}{\textbf{Winter Peak}}  & \multicolumn{3}{c}{\textbf{Light Load}}\\
\midrule
{}   & Island 1 & Island 2 & Island 3 & Island 1 & Island 2 & Island 3 & Island 1 & Island 2  & Island 3 \\
\textbf{Gen. redispatch (MW) }  &  -3341.4 & 1544.6   & 1489.22 & 1418.9 & -4525.2 & 3294.9 & 1313.8 & -2433.4 & 1085.3\\
\textbf{Load-shedding (MW}) & 0 & 0 & 30 &0 & 0 & 0 & 0 & 0 & 0\\
\textbf{Min. bus voltage (p.u.)}  & 0.96 & 0.95 & 0.96 & 0.97 & 0.96 & 0.96 & 0.96 & 0.94 & 0.96 \\
\bottomrule
\end{tabular}
\label{tab:3island}
\vspace{-0.15in}
\end{table*}

\vspace{-0.1in}
\subsection{Discussions}
\vspace{-0.02in}

The islands determined geographically seem to make sense. To measure the quality of partition, we define a metric $p$ that is a ratio of the power flow on tie lines to be disconnected and the power flow within islands. Evidently, a smaller value of $p$ indicates that zones within an island are tightly coupled. The ratio metric has been used to compare performance for different system load conditions. 
\vspace{3pt}
\[p=\frac{\text{Sum of edge weights connecting clusters}}{\text{Sum of edge weights within clusters}}
\vspace{3pt}
\]

 Table \ref{tab:p} shows the metric $p$ for different seasons (calculated before any generation re-dispatch or load shedding). It can be seen that splitting the network requires high power flow disruption during light load conditions. This follows our previous discussion that there is high inter-area power flow during light load to minimize energy prices. However, during light load conditions there will be higher generation reserves that may be redispatched to sustain the islands.

A summary of the electrical performance of the islands created for each case (due to space limitations, we show the results only for the two and three island cases) is provided in tables \ref{tab:2island} and \ref{tab:3island}. These values have been determined using steady state power flow solutions with the PSS/E model. In our simulation, generation has been redispatched and voltage control actions have been performed before splitting the network into islands and solving power flow. It is evident from these values that the islands can be sustained, albeit with operator intervention. The fact that PJM has adequate additional installed generation capacity keeps load-shedding required to a minimum. Of course, the exact sequence of operations for grid splitting remains to be determined.

%% file: conclusion.tex
\vspace{-0.08in}\section{Conclusion}\label{sec:conclusion}


PJM frequently implements operating procedures to posture the system, enhancing reliability during stressed conditions (weather alerts, geomagnetic disturbances, gas-pipeline contingencies, etc). The control actions include procuring more synchronous reserves, decreasing the power transfer across major interfaces, cancelling scheduled outages, etc. In the context of PJM's long-term resilience enhancement goals, 
this paper puts forth a methodology to proactively split the grid  into islands in preparation for HILF events. Strategically islanding the grid has two major benefits: 1) it bounds the extent of potential cascading outages, and 2) it facilitates system restoration.
We also outline several factors that must be considered while deciding how many islands should the network be split into. Our approach is adaptive, independent of fault location, addresses prevailing network conditions and may be tailored according to the nature of the expected contingency. We use the PSS/E model of the heavily meshed PJM transmission network to validate our islanding methodology and check the electrical performance of the resultant islands. 
Realizing the islands in practice would need formulation of an exact sequence of switching operations. This is a complex problem that we intend to pursue in our future research.
\vspace{-0.07in}

%% file: main.bbl
\begin{thebibliography}{10}
\providecommand{\url}[1]{#1}
\csname url@samestyle\endcsname
\providecommand{\newblock}{\relax}
\providecommand{\bibinfo}[2]{#2}
\providecommand{\BIBentrySTDinterwordspacing}{\spaceskip=0pt\relax}
\providecommand{\BIBentryALTinterwordstretchfactor}{4}
\providecommand{\BIBentryALTinterwordspacing}{\spaceskip=\fontdimen2\font plus
\BIBentryALTinterwordstretchfactor\fontdimen3\font minus
  \fontdimen4\font\relax}
\providecommand{\BIBforeignlanguage}[2]{{%
\expandafter\ifx\csname l@#1\endcsname\relax
\typeout{** WARNING: IEEEtran.bst: No hyphenation pattern has been}%
\typeout{** loaded for the language `#1'. Using the pattern for}%
\typeout{** the default language instead.}%
\else
\language=\csname l@#1\endcsname
\fi
#2}}
\providecommand{\BIBdecl}{\relax}
\BIBdecl

\bibitem{HILF}
``{High-Impact, Low-Frequency Event Risk to the North American Bulk Power
  System},'' Department of Energy, Tech. Rep., 06 2010.

\bibitem{weather_CRS}
{Richard J. Campbell}, ``{Weather-Related Power Outages and Electric System
  Resiliency },'' Congressional Research Service, Tech. Rep. R42696, Aug. 2012.

\bibitem{cybersecurity_CRS}
``{Electric Grid Cybersecurity},'' {Congressional Research Service}, Tech. Rep.
  R45312, Sep. 2018.

\bibitem{PJM_manual13}
{Systems Operations Division}, ``{PJM Manual 13: Emergency Operations},'' PJM
  Interconnection, Tech. Rep., Aug. 2019.

\bibitem{meyur}
R.~Meyur, A.~Vullikanti, M.~V. Marathe, A.~Pal, M.~Youssef, and V.~Centeno,
  ``Cascading effects of targeted attacks on the power grid,'' in \emph{Complex
  Networks and Their Applications VII}.\hskip 1em plus 0.5em minus 0.4em\relax
  Springer International Publishing, 2019, pp. 155--167.

\bibitem{cascading_tree}
E.~{Bernabeu}, K.~{Thomas}, and Y.~{Chen}, ``Cascading trees power system
  resiliency,'' in \emph{2018 IEEE/PES Transmission and Distribution Conference
  and Exposition (T D)}, April 2018, pp. 1--9.

\bibitem{Li_controlled_partitioning}
J.~{Li}, C.~{Liu}, and K.~P. {Schneider}, ``Controlled partitioning of a power
  network considering real and reactive power balance,'' \emph{IEEE
  Transactions on Smart Grid}, vol.~1, no.~3, pp. 261--269, Dec 2010.

\bibitem{tortos}
J.~{Quiros-Tortos}, R.~{Sanchez-Garcia}, J.~{Brodzki}, J.~{Bialek}, and
  V.~{Terzija}, ``Constrained spectral clustering-based methodology for
  intentional controlled islanding of large-scale power systems,'' \emph{IET
  Generation, Transmission Distribution}, vol.~9, no.~1, pp. 31--42, 2015.

\bibitem{ICI_measurement}
M.~{Awadalla}, P.~N. {Papadopoulos}, and J.~V. {Milanović}, ``An approach to
  controlled islanding based on {PMU} measurements,'' in \emph{2017 IEEE
  Manchester PowerTech}, June 2017, pp. 1--6.

\bibitem{ICI_MILP}
P.~{Demetriou}, A.~{Kyriacou}, E.~{Kyriakides}, and C.~{Panayiotou}, ``Applying
  exact {MILP} formulation for controlled islanding of power systems,'' in
  \emph{2016 51st International Universities Power Engineering Conference
  (UPEC)}, Sep. 2016, pp. 1--6.

\bibitem{ICI}
A.~{Kyriacou}, P.~{Demetriou}, C.~{Panayiotou}, and E.~{Kyriakides},
  ``Controlled islanding solution for large-scale power systems,'' \emph{IEEE
  Transactions on Power Systems}, vol.~33, no.~2, pp. 1591--1602, March 2018.

\bibitem{ICI_ahad}
A.~{Esmaeilian} and M.~{Kezunovic}, ``Prevention of power grid blackouts using
  intentional islanding scheme,'' \emph{IEEE Transactions on Industry
  Applications}, vol.~53, no.~1, pp. 622--629, Jan 2017.

\bibitem{pjm_value}
``The benefits of the {PJM} transmission system,'' PJM Interconnection, Tech.
  Rep., 04 2010.

\bibitem{panteli_weather}
M.~{Panteli}, D.~N. {Trakas}, P.~{Mancarella}, and N.~D. {Hatziargyriou},
  ``Boosting the power grid resilience to extreme weather events using
  defensive islanding,'' \emph{IEEE Transactions on Smart Grid}, vol.~7, no.~6,
  pp. 2913--2922, Nov 2016.

\bibitem{spectral}
R.~J. {Sanchez-Garcia}, M.~{Fennelly}, S.~{Norris}, N.~{Wright}, G.~{Niblo},
  J.~{Brodzki}, and J.~W. {Bialek}, ``Hierarchical spectral clustering of power
  grids,'' \emph{IEEE Transactions on Power Systems}, vol.~29, no.~5, pp.
  2229--2237, Sep. 2014.

\bibitem{spectral_clustering_tutorial}
U.~{von Luxburg}, ``{A Tutorial on Spectral Clustering},'' \emph{arXiv
  e-prints}, p. arXiv:0711.0189, Nov 2007.

\bibitem{optimal_switching}
K.~W. {Hedman}, S.~S. {Oren}, and R.~P. {O'Neill}, ``A review of transmission
  switching and network topology optimization,'' in \emph{2011 IEEE Power and
  Energy Society General Meeting}, July 2011, pp. 1--7.

\end{thebibliography}
